# Design of Automatic Soil Humidity Control using Maximum Power Point Tracking Controller

Choo Kian Hoe[1], Aravind Chokalingam Vaithlingam[2], Rozita Teymourzadeh[3]  Rajparthiban Rajkumar[4]
[1,2] Energy & Power Research Cluster UCSI University, Kuala Lumpur, Malaysia
[2,3] Faculty of Engineering, Architecture & Built Environment  UCSI University, Kuala Lumpur, Malaysia
[4] Taylors University College, Malaysia
[2] aravind_147@yahoo.com

**Abstract-** The photovoltaic system uses the photovoltaic array as a source of electrical power for the direct conversion of the sun's radiation to direct current without any environmental hazards. The main purpose of this research is to design of a converter with Maximum Power Point Tracker (MPPT) algorithm for any typical application of soil humidity control. Using this setup the major energy from the solar panel is used for the control of soil humidity. The design of the converter with MPPT together with the soil humidity control logic is presented in this paper. Experimental testing of the design controller is implemented and evaluated for performance under laboratory environment.

## I. INTRODUCTION

Water and energy resources are the most essential elements for sustaining human needs, maintaining health and food production, as well as for social and economic development. Based on previous research work, the capital city in the text is endeavouring hard to ascertain ways and means to keep the water level at the optimum level by finding an alternative water supply to maintain the high rising demand for water [1]. The actual cause of this high demand for water is due to the improper usage and handling of the clean water. In order to sustain the rising demand of energy, some alternative energy resources have been proposed in recent years such as solar energy, wind and tidal resources.

Solar energy is prevalent popular choice of clean, recyclable renewable energy as they are more environmental friendly. Unfortunately, the optimal utilisation of the photovoltaic [5-7] energy is a major drawback with the conversion efficiency being in the range of 8 to 16 percent. This percentage is even lower with lesser radiation from the sun due to the improper weather conditions. In order to overcome these problems, the converter is designed with Maximum Power Point Tracking (MPPT) system.

This paper presents a simple photovoltaic rainwater pumping system which is taken as an application to demonstrate the soil humidity control with the MPPT-based converter. The soil humidity controller acts automatically when the humidity of the system drops thereby pumping the rainwater that is collected underneath in the reservoir. The humidity level can be adjusted by the users using a set point value.

## II. SYSTEM OVERVIEW

The system configuration is divided into two parts, buck converter with MPPT algorithm and soil humidity controller. Figure 1 indicates the overall block diagram representation of the system.

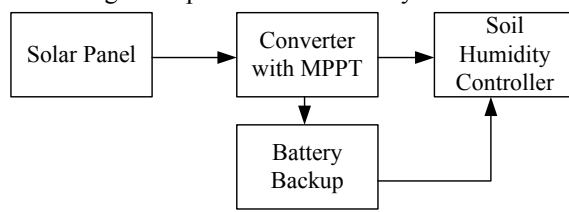

Figure 1: Overview of the Complete System

The buck converter is designed based on explications given in ref [3]. The MPPT algorithm is generated using the PWM where the MOSFET driver activates the corresponding switch operations.

Figure 2 shows the block diagram for the buck converter with MPPT algorithm.

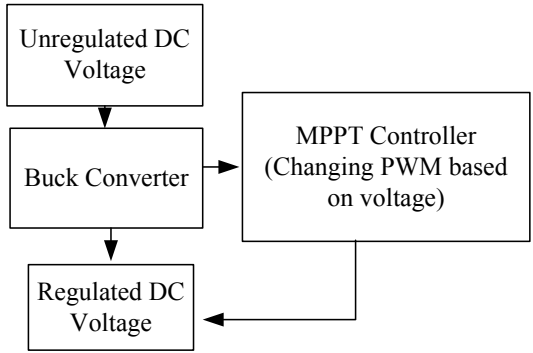

Figure 2: Buck Converter with MPPT algorithm

The soil humidity controller uses a microcontroller to control the soil humidity. Using a humidity sensor such as capacitive sensor, the status of the humidity level is measured and displayed on the LCD screen. The controller activates the external water pump once the humidity level drops to the pre-set value. Figure 4 shows the block diagram for the soil humidity controller. Figure 5 shows the complete schematic for the designed system.

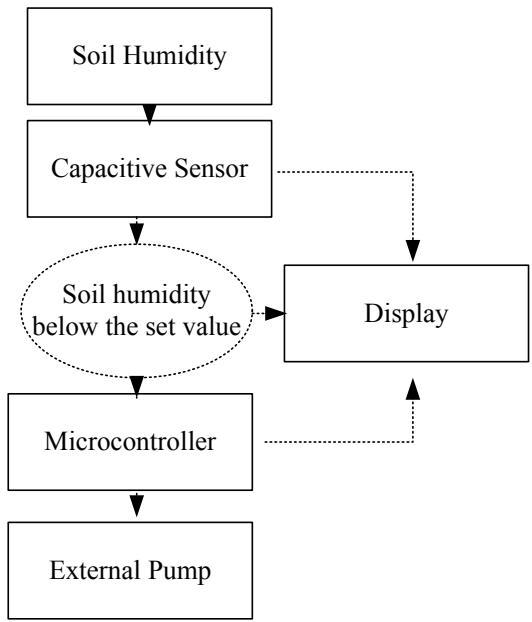

Figure 4: Soil humidity control

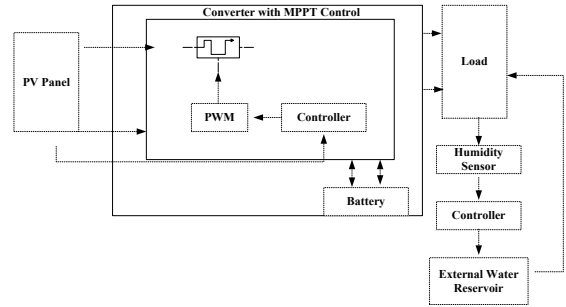

Figure 5: Schematic diagram of the proposed system

### III. CONTROLLER DESIGN

*(a) DC Chopper design*

The DC chopper shown in figure 5 is designed based on the explanations found in reference [3].

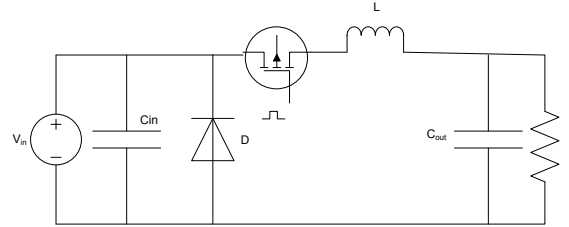

Figure 5: DC-DC converter

The various equations involved in the design are as follows;
The output voltage is given by

$$V_{OUT} = V_{IN}\frac{t_1}{T} = DV_{IN} \quad (1)$$

The design values for L and C are

$$L = \frac{(1-D)R}{2f} \quad (2)$$

$$C = \frac{1-D}{16Lf^2} \quad (3)$$

where
    D : duty cycle
    f : switching frequency
    L : inductor value
    C: capacitor value

The required capacitor value must be greater than this minimum value when designing the converter [8].

*(b) MPPT Controller*

The energy harvested through the photovoltaic array has an optimum point, which is called the Maximum Power Point Tracking (MPPT). This maximum power point cannot be achieved when the voltage or the current of the operating system is incompatible with the solar array [2]. Meanwhile, the maximum power point is also dependent on the temperature and the degree of the solar radiation striking by the panel. Therefore, the Maximum Power Point Tracker method is used to ensure that the solar module to extract the maximum power from the solar panel for rapidly changing conditions to charge the battery. For generating the pulse width, the TL494CN is used instead of PIC or microcontroller since all the functions needed to construct a pulse width modulation is incorporated in this chip. The more flexible TL494 is specially designed for power supply control. The MPPT algorithm can be achieved by using the internal function of the chip. Internally, it has two error amplifiers, an on-chip adjustable oscillator, a Dead Time Control (DTC) comparator, a pulse-steering control flip-flop, a 5V precision regulator and output control circuits. The TL494 achieved the MPPT by using the error amplifiers that measures the output signal and the sets the value to automatically change the PWM of the converter to achieve the maximum point of the power supplied to the converter [4].

*(c) Soil Humidity Controller*

A microcontroller PIC16F887 is used for the soil humidity sensing and controller application. It has the feature of sensing by means of the soil humidity sensor and displays the humidity level on the LCD screen. The controller activates the external water pump once the humidity level drops to the set value. The built-in ADC is suitable to read the soil humidity since the output signal from the humidity sensor is in analog. The humidity sensor is supplied with 5V DC voltage and the output signal is based on the capacitance value of the sensor. Figure 6 shows the ADC value programmed into the controller.

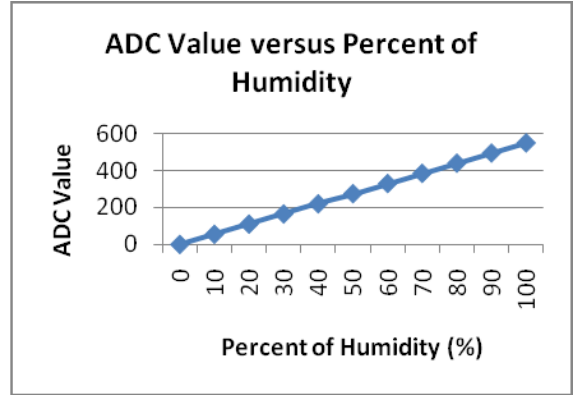

Figure 6: ADC measurements versus Humidity

IV. HARDWARE OVERVIEW

Figure 7 shows the prototype for the experimental setup operating by extracting the maximised energy from the solar panel. The solar panel can be adjusted based on the geographic coordinate of the place, since different place has different coordinate of sun radiation that strikes it at a different angle on the plane of the solar panel. Once the energy is generated, the buck converter with Maximum Power Point Tracker (MPPT) extracts the most energy from the solar panel to use in the soil humidity controller and store some of the energy in the battery for back-up operation. From this, the system performs autonomously with sufficient/adequate sun radiation and rainwater.

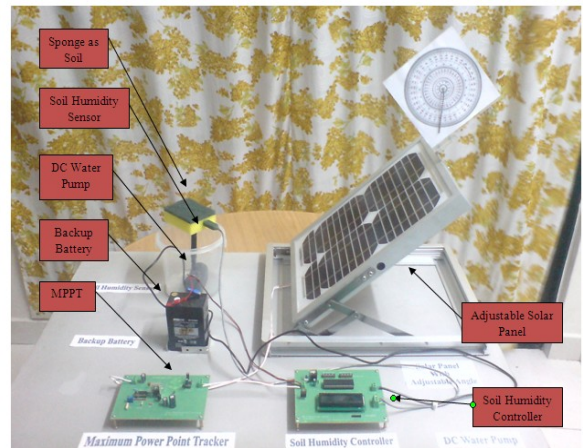

Figure 7: Prototype experimental setup

## V. RESULTS AND DISCUSSION

Figure 8 shows the three different radiation levels on the solar panel at various times of the day to derive the solar characteristics. Figure 9 shows the three different solar radiations on the solar panel based on the power and voltage curves of the solar panel. The PV graph is plotted to show the maximum power point of the solar panel for optimal control. Figure 10 shows the time and voltage characteristics of the converter. The charging point of the converter based on the time of day indicated can be determined from this plot.

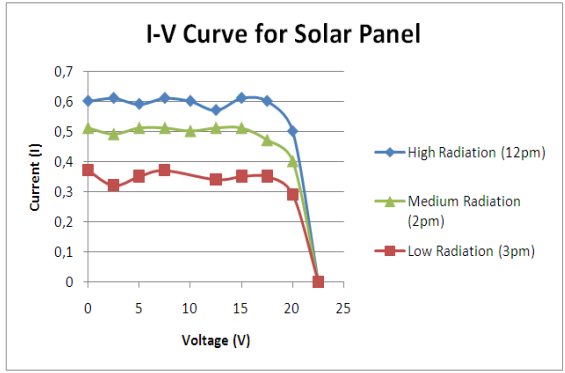

Figure 8: Solar panel radiation characteristics

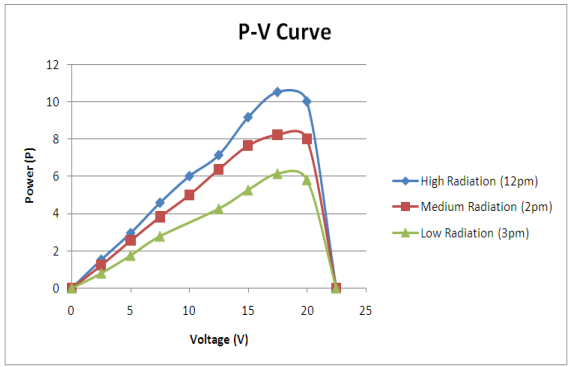

Figure 9: P-V Curve of the Solar Panel

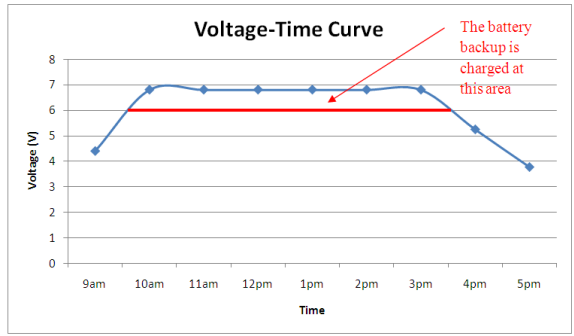

Figure 10: Converter voltage characteristics

Figure 11 shows the time and current value for the converter. From this value, the Current-Time Curve graph is plotted to obtain the charging current of the converter at different time. Figure 12 shows the Power-Time Curve graph to show the maximum power of the converter

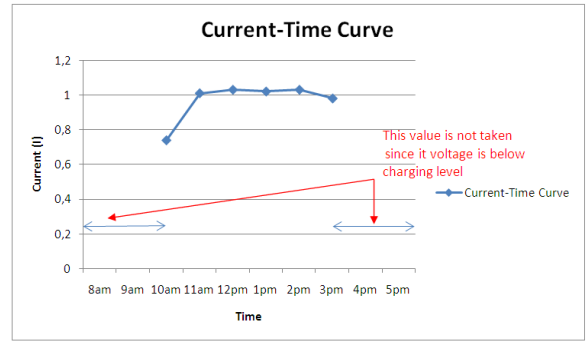

Figure 11: Converter current characteristics

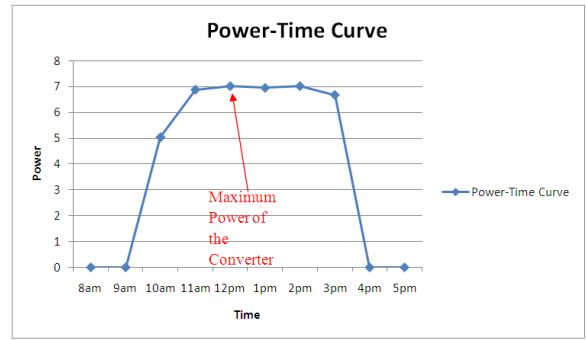

Figure 12: Power-Time Curve of the Converter

## VI. CONCLUSIONS

The converter with MPPT algorithm is designed constructed, built and validated for performance. This power output is able to be used in soil humidity controller with battery backup. The external source is able to supply water to the soil when the humidity level is dropped to the desired humidity level. This research has a contributory significance to the current endeavours for safe, renewable and environmental friendly alternative source of energy in the face of countering the adverse human intervention effects on the global climatic change environment.